\documentstyle[epsf,prb,preprint,aps]{revtex}
\tightenlines

\newcommand{\bA}{\mbox{\boldmath$A$}}     
\newcommand{\bef}{\mbox{\boldmath$f$}}     
\newcommand{\bk}{\mbox{\boldmath$k$}}     
\newcommand{\bv}{\mbox{\boldmath$v$}}     
\newcommand{\bS}{\mbox{\boldmath$S$}}

\begin{document}

\draft
\title{Coexistence of spin-triplet superconductivity and ferromagnetism
induced by the local Hund's rule exchange}

\author{Jozef Spa\l ek and Pawe\l\ Wr\'{o}bel}
\address{Marian Smoluchowski Institute of Physics, 
Jagiellonian University, \\
ulica Reymonta 4, 30-059 Krak\.ow, Poland}

\date{\today}
\maketitle

\begin{abstract}

We characterize the coexistence of itinerant ferromagnetism and spin-triplet
superconductivity within a single mechanism involving local (Hund's rule) exchange
among $d$ electrons. The ratio of transition temperatures and the spin
anisotropy of the superconducting gap is estimated for $\rm ZrZn_{2}$.
The $A$ phase is stable in very low applied and molecular fields, whereas
the $A1$ phases persists in higher fields. A small residual magnetic moment
is present below the Stoner threshold in the superconducting phase.

\end{abstract}

\pacs{PACS}
\newpage

The coexistence of weak itinerant ferromagnetism (WFM) and
superconductivity (SC) has been recently discovered in $\rm UGe_{2}$
(Ref.~1), $\rm ZrZn_{2}$ (Ref.~2) and $\rm URhGe$ (Ref.~3). The
superconducting phase was encountered {\it in the ferromagnetic phase};
both of these phases seem to disappear with increasing pressure. Therefore,
the superconductivity must be influenced by the ferromagnetism,
particularly since the ratio of the Curie temperature $(T_{c})$
relative to the superconducting transition temperature $(T_{s})$ can exceed
an order of magnitude. 

It is  hard to imagine that the
superconducting pairing in that situation involves a spin singlet, since the
molecular field,  e.g., in $\rm ZrZn_{2}$ due to the exchange interaction,
which is of the order of \hspace{1ex}\cite{koelling71} 
$H_{m}$ =  150 T $(\mu_{B} H_{m}$ = 17 meV),   exceeds by far the thermodynamic critical field $H_{c} \leq
1\,T$.  

In this paper we consider
both of these types of ordering within a single mechanism -- the Hund's
rule exchange and we draw some universal conclusions from a
relatively simple and testable model containing two microscopic parameters
(apart from the density of states and its derivatives at the Fermi level).
We follow some of the approximations of the original Bardeen,
Cooper, and Schrieffer (BCS) theory, although we employ the linearized
dispersion relation and spin-split structure of quasiparticle states to
account for the Fermi-liquid structure of the weak-ferromagnetic such as
$\rm ZrZn_{2}$.

The $\it {spin-triplet}$ pairing in weakly ferromagnetic systems has been considered
before being mediated by the exchange of longitudinal spin
fluctuations \cite{fay80}  or being triggered by the electron-phonon 
interaction.\cite{enz78}\hspace{2ex}Quite recently 
\cite{karchev01,larkin64}, the question of
the coexistence of ferromagnetism with $\it {spin-singlet}$ superconductivity has
been reexamined within the mean-field theory. All the foregoing work is
based on one-band (Hubbard or extended Hubbard) model, so the
superconductivity can arise either from exchange of a paramagnon for
repulsive interactions \cite{fay80}, or as a result of local attractive
interaction ($\it negative-U \; model$).\cite{karchev01} In our approach the pairing, 
induced by local Hund's' rule exchange, appears in the correlated and orbitally
degenerate systems and together with the short-range Coulomb
interaction is regarded as the source of itinerant magnetism in 3d and 4d metallic
system. $\,$\cite{ten}This interorbital interaction remains local if the
degenerate bands are not strongly hybridized. This is exactly what happens
for $\rm ZrZn_{2}$, where  the main contribution to the high density of
states $\rho (\epsilon_{F})$ at the Fermi energy from the two $d$
bands, which do not hybridize \cite{koelling71,johnson74} (see also the
discussion below.) 

We start from the effective model proposed by us recently \cite{spalek01}
and extend it to consider explicitly both ferromagnetism and
superconductivity. It is represented by the Hamiltonian
\begin{equation}
{\cal H} =
\sum_{\bk \ell \sigma}
(E_{\bk \ell} - \mu)
n_{\bk \ell \sigma} +
{\cal H}_{ex} + {\cal H}_{c}
\end{equation}  
The quasiparticle band energy $E_{\bk \ell} \equiv E_{\bk \ell} - \mu$
of quasimomentum $\bk$ is labelled by the orbital index $\ell = 1$ and 2.
The local interorbital and intraatomic Hund's' rule coupling can be represented
in two equivalent ways as \cite{klejnberg99}
\begin{equation}
{\cal H}_{ex} =
-2J \sum_{i}
(\bS_{i1} \cdot \bS_{i2} +
\frac{3}{4} n_{1i}n_{2i}) =
-2J \sum_{im}
A_{im}^{+}
A_{im} ,
\end{equation}
where $\bS_{i\ell}$ and $n_{i\ell}$ are the spin and the particle-number
operators for $\ell$-th orbital on site $i$, and $a_{im}^{+}$ are
real-space spin-triplet creation operators on site $i$ $(A_{i1}^{+} =
a_{i1\uparrow}^{+} a_{2i\uparrow}^{+}$, etc.). ${\cal H}_{c}$ represents
the direct Coulomb interaction (the Hubbard term), as well as the
interorbital Coulomb term
\begin{equation}
{\cal H}_{c} = U \sum_{li} n_{li\uparrow} n_{li\downarrow} +
U' \sum_{l \neq l',i} n_{li} n_{li} \; .
\end{equation}
The last term does not influence the phases considered in the first order 
so it is dropped out.
So, (1) represents, in our
view, a correlated system \cite{large}, in which the local interactions
determine the quantum instabilities.

We introduce the combined Hartree-Fock-BCS approximation and the four
dimensional Nambu-type  representation. This means that the interaction
part is rewritten first (up to a constant) in the following manner
\begin{eqnarray}
{\cal H}_{ex} + {\cal H}_{c}&=& -2J \sum_{m} 
\left(<A_{im}> A_{im}^{\dagger} +
<A_{im}^{\dagger}> A_{im} -
|<A_{im}>|^2 \right) \nonumber \\ 
&& - I \left[ \sum_{i}
\overline{S^{z}}
(S_{i1}^{z} +
S_{i2}^{z}) -
\frac{1}{2} (\overline{S^{z}})^{2} \right] ,
\end{eqnarray}
where $\overline{S^{z}} \equiv < S_{i1}^{z} + S_{i2}^{z} >$ is the
magnetic moment per atom, $<A_{im}>$ is one of the
three possible components of the superconducting gap parameters,
and $I=U+2J$ is the effective Stoner parameter.
Subsequently, the Hamiltonian can be rewritten as a $ 4 \times 4$
matrix with creation operators ${\bf f}_{\bk}^{+} \equiv (f_{\bk 1 \uparrow}^{\dagger}
f_{\bk 1 \downarrow}^{\dagger}, f_{-\bk2\uparrow}, f_{-\bk 2 \downarrow})$, so
that
%eq4
\begin{eqnarray}
{\cal H} &=&
\sum_{\bk} 
\bef_{\bk}^{\dagger}
\bA \bef_{\bk} +
\sum_{\bk \sigma}
E_{\bk 2} + N
\left[
I(\overline{S^{z}})^{2} +
\sum_{m}
\frac{|<A_{im}>|^{2}}{2J} 
\right] ,
\end{eqnarray}
where $\bA$ is $4 \times 4$ matrix of the form
\begin{eqnarray*}
\bA = \left(
\begin{array}{ccccccc}
E_{\bk1} - I\overline{S^{z}} &,& 0 &,& \Delta_{1} &,& \Delta_{0} \\
0 &,& E_{\bk1} + I\overline{S^{z}} &,& \Delta_{0} &,& \Delta_{-1} \\
\Delta_{1} &,& \Delta_{0}  &,& -E_{\bk2} + I\overline{S^{z}} &,& 0 \\
\Delta_{0} &,& \Delta_{-1} &,& 0  &,& -E_{\bk2}-I\overline{S^{z}} 
\end{array}
\right) .
\end{eqnarray*}
This matrix can be diagonalized analytically. The interesting cases are only
those with the spin dependend gaps
(i) $\Delta_{\uparrow} > \Delta_{\downarrow}$ and $\Delta_{0} = 0$ (which
we call the anisotropic $A$ phase); and (ii) $\Delta_{\uparrow} \neq 0$ and
$\Delta_{\downarrow} = \Delta_{0} = 0$ (which is called the $A1$ phase). In case
(i), the four eigenvalues are
%eq5
\begin{equation}
\lambda_{\bk \sigma 1,2} =
\frac{1}{2} (E_{\bk 1} - E_{\bk 2}) \mp
\left[ \frac{1}{4} (E_{\bk 1} + 
E_{\bk2} - 
\sigma I \overline{S^{z}})^{2} +
\Delta_{\sigma}^{2} \right]^{1/2} .
\end{equation} 
The sign $(\mp)$ corresponds to the labels (1,2) of $\lambda_{\bk \sigma}$
and reflects hole and electron excitations, respectively. The spectrum is separated with
respect to the spin orientation $\sigma = \pm 1$ of the Cooper pair. The
spectrum is fully gapped if both $\Delta_{\uparrow}$ and
$\Delta_{\downarrow}$ are nonzero. We obtain a combination of the spin
splitting and superconducting gap (the gap at the Fermi energy $E_{F} =
\mu$) is
$2 \left( \sqrt{(I\overline{S^{z}} )^{2} + 
\Delta_{\uparrow}^{2}} +
\sqrt{(I\overline{S^{z}})^{2} + 
\Delta_{\downarrow}^{2}} 
\right) \sim
2I\overline{S^{z}}$ 
when the spin is flipped and $\Delta_{\uparrow}$ or
$\Delta_{\downarrow}$ within the spin subband. What is probably more
important, in $A1$ state half of the spectrum remains gapless 
$\lambda_{\bk\downarrow 1} = - E_{\bk 2} + I\overline{S^{z}}$,
$\lambda_{k\downarrow 2} = E_{\bk 1} + I\overline{S^{z}}$. Thus, there should
be a {\it substantial linear specific-heat term present} also in the
superconducting $A1$ state and this dependence should be distinguished from the
$T^{n}$ dependence $(n \geq 2)$ due to the  gap zeros (the latter would
require the interband hybridization and hence the dependence \cite{jsunpub}
$J \rightarrow J_{\bk\bk'})$. 
The linear specific heat $\gamma$ comes
from the spin minority electrons, and since for a
weak itinerant magnet $I\overline{S^{z}} \ll E_{F}$, there will be 
drop in the relative value of $\gamma$ of the order $\rho_{\downarrow} /
(\rho_{\uparrow} + \rho_{\downarrow})$, where $\rho_{\sigma} \equiv
\rho_{\sigma} (E_{F})$ is the density of states in spin-$\sigma$ subband at
the Fermi energy.

The results obtained so far are general in the sense they are independent
of a particular electronic structure. In the following we assume that the bands
are the same, i.e. $E_{\bk 1} = E_{\bk 2} \equiv E_{\bk}$ so that
$E_{\bk} - \mu \simeq {\bv}_{F} \bk$, where $v_{F}$ is the Fermi velocity.

The Bogolyubov quasiparticle operators can also be easily calculated; then they
are \cite{assume}
%eq6
\begin{eqnarray}  
\left( 
\begin{array}{cc}
\alpha_{\bk \sigma} \\
\beta_{-\bk \sigma}^{\dagger}
\end{array}
\right)
= \frac{1}{\sqrt{2}}
\left(
\begin{array}{ccc}
u_{\bk}^{(\sigma)} &,& v_{\bk}^{(\sigma)} \\
-v_{\bk}^{(\sigma)} &,& u_{\bk}^{(\sigma)}
\end{array}
\right)
\left(
\begin{array}{c}
f_{\bk 1 \sigma} + f_{-\bk 2 \sigma}^{\dagger} \\
f_{\bk 1 \sigma} - f_{-\bk 2 \sigma}^{\dagger}
\end{array}
\right) ,
\end{eqnarray}
with the coherence factors
%eq7
\begin{eqnarray}
\left(
\begin{array}{c}
u_{\bk}^{(\sigma)} \\
v_{\bk}^{(\sigma)}
\end{array}
\right)
= \frac{1}{\sqrt{2}}
\left(
\begin{array}{ll}
1 + \Delta_{\sigma} / \lambda_{\bk \sigma} \\
1 - \Delta_{\sigma} / \lambda_{\bk \sigma}
\end{array}
\right) ,
\end{eqnarray}
and the eigenvalue $\lambda_{\bk \sigma} = \lambda_{\bk \sigma 1} =
- \lambda_{\bk \sigma 2} \equiv
[(v_{\sigma}k)^{2} + \Delta_{\sigma}^{2} ] ^{1/2}$. 
Note that the spin dependent Fermi velocity is caused by the circumstance that
we have spin-split bands in the ferromagnetic phase.
The Hamiltonian has the diagonal form
${\cal H} = \sum_{\bk \sigma} \lambda_{k\sigma} (\alpha_{\bk
\sigma}^{+}\alpha_{\bk \sigma} + \beta_{-\bk \sigma} \beta_{-\bk \sigma}) +
E_{0}$, which is the BCS form with spin-dependent quasiparticle and gap
energies. The last factor contributes to important differences with the BCS
theory. Namely, the gap parameter at temperature $T = 0$ is determined from
the equation 
\begin{eqnarray}
1 = J \rho_{\sigma}
\int_{-k_{m \sigma}}^{k_{m\sigma}}
\frac{d^{3} (v_{\sigma} \bk)}
     {\sqrt{(v_{\sigma} k)^{2} + \Delta_{\sigma}^{2}}} ,
\end{eqnarray}
where $\rho_{\sigma} = 12 \epsilon_{F\sigma}^{2} / W$ $(\epsilon_{f\sigma}$
is the Fermi energy for quasiparticles  with spin $\sigma$ and $W$ is the
effective width of the band, related
to the Fermi velocity via $W = (24 \pi^{2} \hbar v_{F}^{2} /
\Omega_{0})^{1/3}$, where $\Omega_{0}$ is the elementary-cell volume).  The
integration boundary $k_{m}$ is determined by the condition that the
paired particles with spin $\sigma$ are present only within the spin-split
region of the band, i.e., by the constraint $ v_{\sigma} k_{m\sigma} = \sigma I
\overline{S^{z}}$. In effect, we
obtain an estimate
%eq9
\begin{equation}
\Delta_{\sigma} = 4 I \overline{S^{z}}
\exp \left(-
\frac{1}{\rho_{\sigma} J}
\right).
\end{equation} 
Analogously, we can estimate the critical temperature by selecting an
equivalent but slightly different representation of the quasiparticle
energies: $E_{\bk} - \mu - \sigma I\overline{S^{z}} = \hbar v_{F} k - \sigma
I\overline{S^{z}}$, where $v_{F}$ is the Fermi velocity in the paramagnetic
phase. Then the condition for $T_{S}$ (for $\Delta = \Delta_{\uparrow})$
reduces to
%eq10
\begin{eqnarray}
1 = I 
\int_{-k_{min}}^{k_{min}}
\frac{ \tanh \left(
         \frac{\hbar v_{F}k - 2J\overline{S^{z}}}
              {2k_{B}T_{S}} \right)} 
     {\hbar v_{F}k-J\overline{S^{z}}}.
\end{eqnarray}
In effect, we obtain the estimate
%eq11
\begin{equation}
k_{B}T_{S} \simeq
2.26 I \overline{S^{z}}
\exp \left( -
\frac{1}{ J\rho}
\right) .
\end{equation}
In both expressions for $\Delta_{\sigma}$ and for $T_{S}$, the exponent
contains the exchange integral, whereas the preexponential factor is
multiplied by the magnetic moment. This means that spin-triplet
superconductivity {\it disappears together} with ferromagnetism. This
result is more general as it relies on the well founded notion that Cooper
per of spin orientation $\sigma$ exist within the corresponding spin
subbands if only $\Delta_{0} = 0$ (the B phase is not stable).

To make the estimates explicit we have to relate the results for a weak
itinerant ferromagnet. $\,$\cite{wohlfarth68}  It is easy  to rederive those
results in the present situation with $T_{c} \gg T_{S}$.  Namely, the Curie
temperature is given by the expression
%eq12
\begin{equation}
T_{c} =
\frac{\sqrt{6}}{\pi}
\left[
\left(
\frac{\rho^{'}}{\rho}
\right)^{2} -
\frac{\rho ''}{\rho}
\right]^{-1/2}
[(I\rho-1)/I\rho]^{1/2},
\end{equation}
with $\rho '$ and $\rho ''$ being, respectively, the first and the second
derivative of the density of states $\rho (E)$ taken at $E_{F}$.
Additionally, the magnetic moment at $T = 0$ is given by
$\overline{S^{z}} = (1/2) [ (I \rho - 1) / B]^{1/2}$, with 
%eq13
\begin{equation}
B =
I^{\,3} \frac{\rho}{8}
\left[ \left( \frac{\rho '}{\rho}\right)^{2} -
\frac{\rho ''}{\rho} \right].
\end{equation}
Taking the parabolic density of states corresponding to the linearized
dispersion relation, we can determine the ratio $T_{s}/T_{c}$ in an explicit
manner
%eq14
\begin{equation}
\frac{T_{S}}{T_{c}} \simeq
1.17 \exp
\left( -
\frac{1}{J\rho}
\right).
\end{equation}
Similarly, the gap ratio is
\begin{equation}
\frac{\Delta_{\uparrow}}{\Delta_{\downarrow}} \simeq
\exp \left(
(2I/3J)(\rho'/\rho^{2})
\overline{S^{z}}
\right),
\end{equation}
which in the limit $I\rho \approx 1$ gives the ratio exceeding three.
The ratio will
grow rapidly with $\overline{S^{z}}$ and the prediction $\ln (
\Delta_{\uparrow} / \Delta_{\downarrow}) \sim \overline{S_{z}}$ could be
tested experimentally. The $A1$
superconducting state is reached fast with the increasing moment.

To interpret our findings in the coexistence regime we can say that the $A$
phase (with $\Delta_{\downarrow} \neq 0)$ cannot appear either near the
quantum critical point -- the Stoner threshold, at which $I \rho = 1$, or
in the ferromagnetically saturated state.  In the former case, the
relatively small field can polarize totally the system (since the magnetic
susceptibility $\chi = \chi_{0} / (1 - I \rho )$ is almost divergent).
Under these circumstances, the bound state with $\Delta_{\downarrow} \neq 0$ 
cannot be formed, since the polarized surrounding will respond to the presence of 
the second electron strongly. On the  other hand, if the magnetic moment in the
ferromagnetic state is almost saturated, then the bound state 
with $\Delta_{\downarrow}$ cannot be
formed as the system is rigid. 
Only in between those two limiting situations, but rather on the
weak-ferromagnetic side, the coherent $A$ state can be realized.
Otherwise, the $A1$ state will become stable. 

To gain a quantitative insight into the nature of $A$ and $A1$ states and
their coexistence with the spin polarized state, we have also performed the
analysis in the paramagnetic state, i.e., for $ \rho I  < 1$, but in the
applied magnetic field $B \neq 0$. To amplify numerically the effects discussed,we have put $I = J $ and choose a  constant density of states 
.  The phase diagram as a function of
magnetic field is displayed in Fig.~1.  Note again that the phase $A$
disappears in the vicinity of the Stoner threshold marked by the vertical
dashed line. On the contrary, the phase $A1$ continues towards the Stoner
boundary.  In Fig.~2, we provide the field
dependence of the  ground state energy away from the Stoner threshold for normal,
$A$, and $A1$ phases. 
The $A1$ phase is the most stable in a high applied field 
and can coexist with a saturated ferromagnetism.  The same
happens for the superfluid $\rm {}^{3}He$. This means that the superconducting
coherence length for $A1$ phase becomes unbound when the system reaches the saturated
state.

The interesting feature of our results is that the presence of the
superconductivity indices a magnetic moment even below the Stoner
threshold. The magnitude of the moment below the Stoner threshold $(J/W =
0.25$ in this case) is displayed in Fig.~3. Obviously, the effect is
enhanced by the choice of the density of states selected and particularly 
by the assumption $I = J$
, but it is worth mentioning. It suggests that in the presence of the superconducting
state makes the Stoner critical point a {\it hidden} one. This point will be elaborated
further elsewhere.

In summary, we have considered the most natural model for the coexistence
of ferromagnetism and spin-triplet superconductivity, both induced by a
single mechanism -- the local ferromagnetic exchange in the orbitally degenerate
systems. While the anisotropic $A$-state seems to be stable in weakly
polarized and paramagnetic systems at low fields, the $A1$ state may coexist
even closer  to the quantum critical point. The paired state induces a small 
magnetic moment even below the Stoner threshold. It should be interesting
to extend these results to the hybridized systems \cite{spalek01}, when
$\bk$-dependent  (p- or d-wave) spin-triplet superconductivity will appear.

\vspace*{12pt}
This work was supported by the KBN Grant No. 2P03B 092 18, as well as by
NSF Grant No. DMR 96-12130.

\newpage

{\centerline{\large {\bf Figure Captions}}} 

\vskip1.0cm

Fig.1. Phase diagram specifying the stable superconducting states A and A1  
(see main text). The dotted line specifies the onset of the magnetically
saturated state. The vertical dot-dashed line marks the Stoner threshold. 

\vskip1.0cm

Fig.2. Ground-state versus applied magnetic field B. At $B = 0$ the
equal-spin pairing state (A-phase) is stable, whereas in high field A1
phase (with the spins parallel) prevails. The system is below the Stoner
threshold.

\vskip1.0cm

Fig.3. Magnetic moment per orbital (upper panel) and the superconducting gaps
(lower panel) 
{\it versus} applied field (below the Stoner threshold). The inset
shows the chemical potential vs. $B$. The horizontal arrow indicates a small
residual magnetic moment in the $B \rightarrow 0$ limit below the Stoner
threshold.

\newpage
\begin{figure}
\epsfxsize15cm
\epsffile{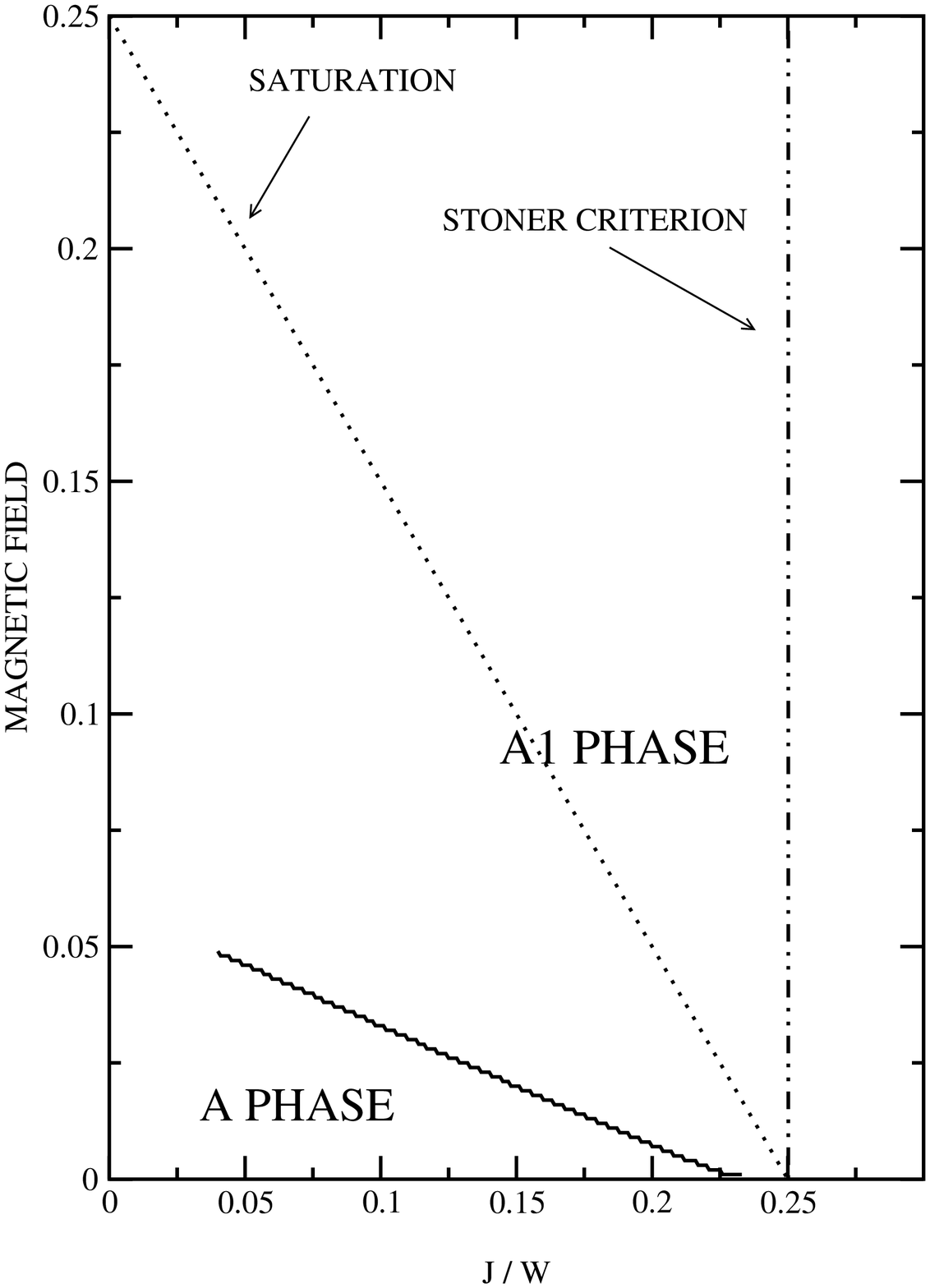}
\caption{}
\end{figure}

\newpage
\begin{figure}
\epsfxsize15cm
\epsffile{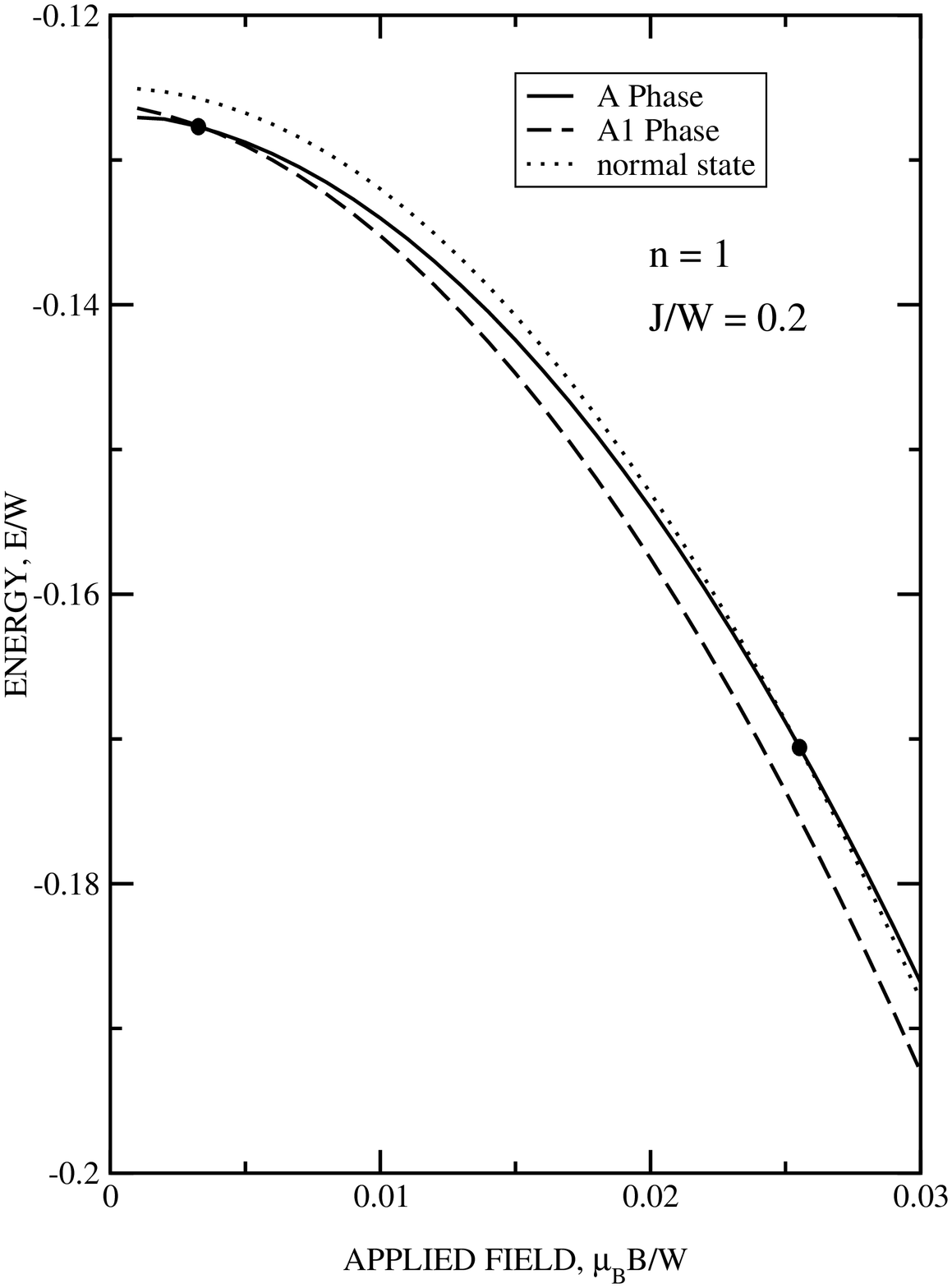}
\caption{}
\end{figure}

\newpage
\begin{figure}
\epsfxsize15cm
\epsffile{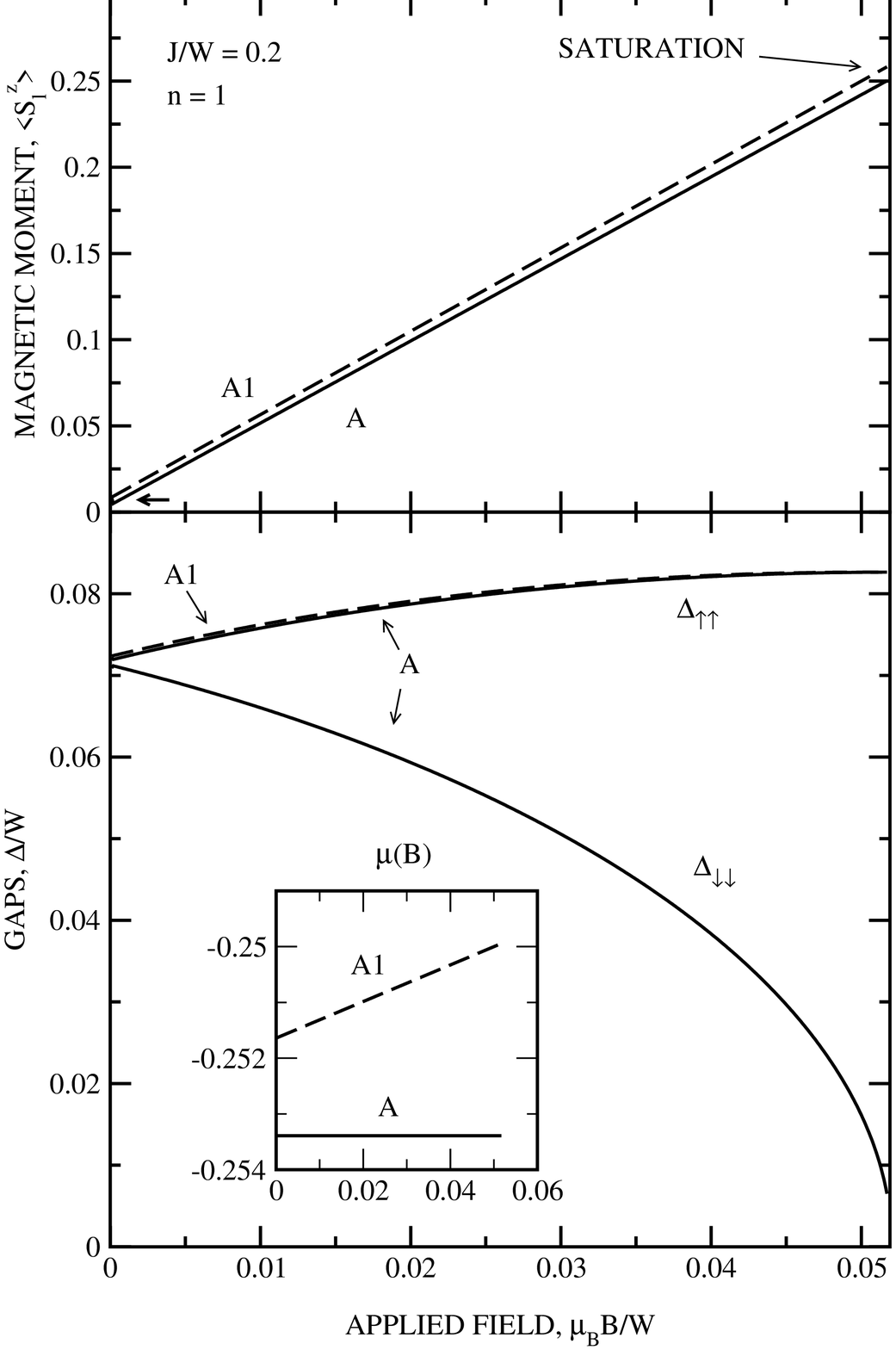}
\caption{}
\end{figure}


\begin{references}
\bibitem{assume}
We assume here that the Fermi velocities are the same in both bands (flat
bands$^{11}$), but they will depend on the spin so that we can represent
the spin-split bands near $E_{F}$ in the form:
$\lambda_{k\sigma} \simeq \hbar v_{F\sigma} (k-k_{F}) \equiv v_{\sigma}k$.
\bibitem{details}
The details of density of states should not matter for qualitative features
of the results, as we integrate now over the whole band states.
\end{references}
\end{document}